# GPU Accelerated Explicit Time Integration Methods for Electro-Quasistatic Fields

Christian Richter[1], Sebastian Schöps[2] and Markus Clemens[1], *Senior Member, IEEE*

[1]Chair of Electromagnetic Theory, University of Wuppertal, 42119 Wuppertal, Germany, *christian.richter@uni-wuppertal.de*
[2]Graduate School of Computational Engineering and Institut für Theorie Elektromagnetischer Felder, Technische Universität Darmstadt, 64285 Darmstadt, Germany, *schoeps@gsc.tu-darmstadt.de*

**Electro-quasistatic field problems involving nonlinear materials are commonly discretized in space using finite elements. In this paper, it is proposed to solve the resulting system of ordinary differential equations by an explicit Runge-Kutta-Chebyshev time-integration scheme. This mitigates the need for Newton-Raphson iterations, as they are necessary within fully implicit time integration schemes. However, the electro-quasistatic system of ordinary differential equations has a Laplace-type mass matrix such that parts of the explicit time-integration scheme remain implicit. An iterative solver with constant preconditioner is shown to efficiently solve the resulting multiple right-hand side problem. This approach allows an efficient parallel implementation on a system featuring multiple graphic processing units.**

***Index Terms*—algebraic multigrid, electro-quasistatic, parallelism, GPUs, explicit time-integration**

## I. Introduction

ELECTRICAL field grading using microvaristor materials embedded in polymeric materials is a field of ongoing research for high-voltage applications as for example insulators, bushings and surge arrestors. The materials change their electrical conductivity depending on the local electrical field with it rising by several orders of magnitude at a switching point. For proper use and design complex 3D models as shown in Fig. 1 must be numerically analyzed.

The electro-quasistatic (EQS) approximation of Maxwell's equation is applied to simultaneously consider capacitive and resistive effects. For solving these problems, they must be discretized in space and time. Space discretization is typically carried out by the Finite Element Method (FEM) and time discretization by sequential time-integration schemes. The most common approaches are based on implicit time integration schemes, like the implicit Euler scheme or the Singly Diagonal Implicit Runge-Kutta (SDIRK) method [1,2]. To solve the nonlinear problem in each time step, an iterative linearization method, as e.g. the Newton-Raphson scheme, is applied and thus many linear algebraic systems need to be solved [3].
The time-consuming computation of large-scale models often exceeds multiple days. However, it can be accelerated by using graphic processor units (GPUs) [4]. Particularly, iterative linear solvers based on sparse matrix-vector operations highly benefit from GPUs [5]. For example, algebraic multigrid (AMG) preconditioners are often used [6] due to their (almost) optimal asymptotic complexity. On the other hand, already medium sized FEM models hit the limits of a contemporary single GPU's global memory. Therefore, multi-GPU AMG-preconditioned conjugate gradients (AMG-CG) solvers have been presented [7,8]. These solvers allow fast solutions of the linear equation systems, but the repeated construction of the preconditioner for the Jacobian and its upload to the GPUs is still a bottleneck of these schemes.

Corresponding author: C. Richter (e-mail: christian.richter@uni-wuppertal.de).

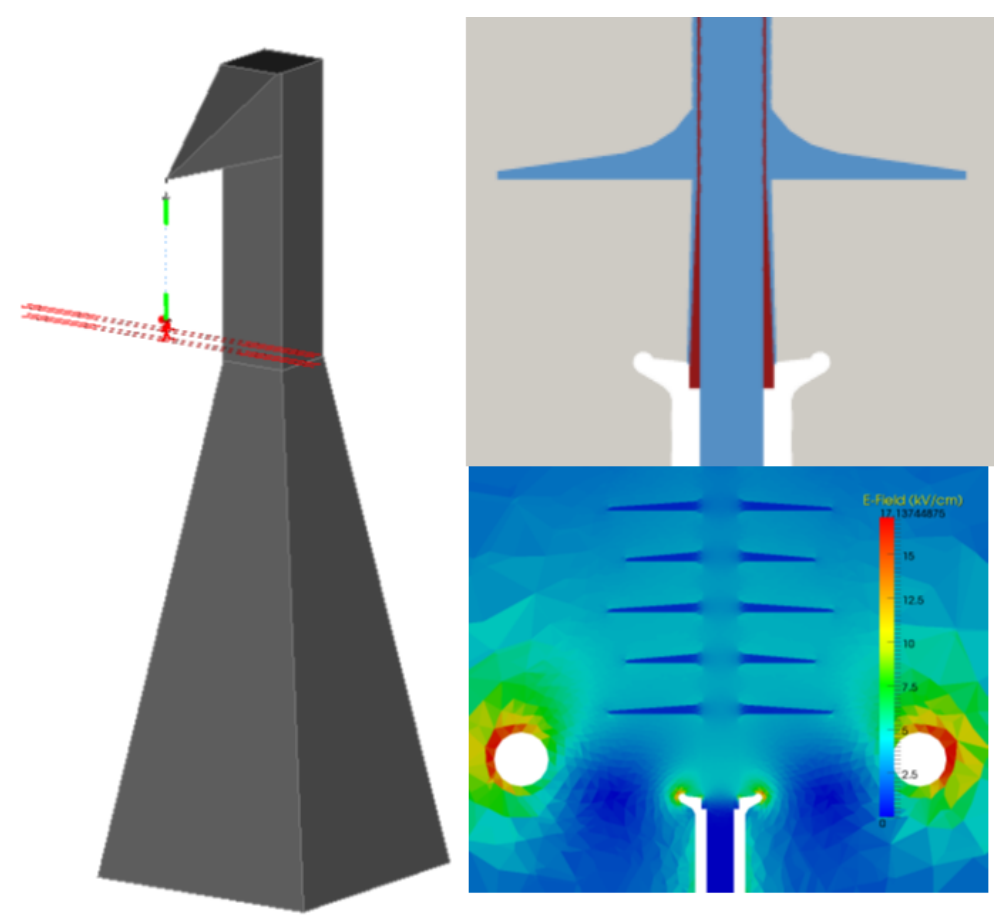

Fig. 1: CAD model and electrical field of the calculated model involving nonlinear microvaristor materials, shown in red on the upper left.

Therefore, this paper deals with the application of explicit time-integration schemes such as the explicit Euler method or the more sophisticated Runge-Kutta-Chebyshev method [1,9]. In the case of EQS, the resulting scheme is not entirely explicit since the mass matrix is a discretization of the electrostatic Laplacian operator. Nonetheless, due to constant permittivity, the explicit scheme leads to a multiple right-hand side (MRHS) problem. This favors the parallelization of the linear algebra since communication costs required for the (nonlinear) Jacobian matrix updates are avoided. Especially the acceleration by GPUs compensates for the increased effort due the time-step stability restriction of the explicit schemes.

The paper is structured as follows: Section II introduces the formulation of the EQS problem and an explicit time-integration scheme is described. A GPU-based combined matrix assembly and sparse matrix vector multiplication is discussed in Section III. Efficient approaches for the MRHS problem are described in Section IV. The methodology is validated by numerical examples in Section V and the paper closes with conclusions in Section VI.

## II. Explicit Time-integration for EQS

The EQS approximation of Maxwell's equations can be formulated with the following parabolic problem

$$\operatorname{div}\left(\kappa(\phi)\operatorname{grad}\phi\right)+\operatorname{div}\left(\varepsilon\operatorname{grad}\frac{\partial\phi}{\partial t}\right)=0, \tag{1}$$

where $\phi$ is the electrical potential, $\varepsilon$ is the permittivity, $\kappa$ is the nonlinear conductivity which depends on the electric field intensity represented by the gradient of electrical potential [10]. The problem is well posed if adequate initial and (Dirichlet) boundary conditions are specified. Typically, the system is discretized in space using the FEM leading to a system of ordinary differential equations (ODE)

$$\mathbf{M}\frac{\mathrm{d}\mathbf{x}}{\mathrm{d}t}+\mathbf{K}(\mathbf{x})\mathbf{x}=\mathbf{b} \tag{2}$$

with $\mathbf{M}$ being the mass matrix, $\mathbf{x}$ is the vector of electric potentials, $\mathbf{K}(\mathbf{x})$ the stiffness matrix depending on the potentials and $\mathbf{b}$ the contribution of inhomogeneous boundary conditions.

In contrast to the commonly applied implicit time-integration schemes, we propose the usage of explicit time-integration schemes. In the simplest case of the explicit Euler time-integration method, the time derivative is discretized by the forward difference quotient, i.e.,

$$\mathbf{x}^{n+1}:=\mathbf{x}^{n}+\Delta t\cdot\mathbf{M}^{-1}\left(\mathbf{b}(t^{n+1})\text{-}\mathbf{K}(\mathbf{x}^{n})\mathbf{x}^{n}\right), \tag{3}$$

where $\mathbf{x}^{i}$ approximates $\mathbf{x}(t^{i})$ and $\Delta t$ is the time step width. This approach comes with advantages and disadvantages compared to implicit time-integration schemes: most importantly, there is a restriction on the time-step size due to the CFL stability condition [1]. This may lead to very small maximal time step size for stiff problems. To this end, the Runge-Kutta-Chebychev (RKC) scheme has been proposed [9]. This explicit scheme adds a variable number of stages during the solution of one time step to extend the stability region quadratically. It was formulated for parabolic problems of heat equation-type with a diagonal mass matrix. Here, we extend the scheme to the non-diagonal case but are faced with additional costs due to the inversion of the mass matrix $\mathbf{M}$.

However, the mass matrix $\mathbf{M}$ is constant and the linear systems to be solved in each time step pose a multiple right-hand side (MRHS) problem. Thus, once the preconditioner is set up, it can be reused in every time-step. Secondly, the stiffness matrix that must be assembled in every Newton iteration in implicit schemes is no longer part of the system matrix. It is a vector resulting from a matrix-vector product well suited for GPUs acceleration.

## III. GPU Accelerated FEM Assembly

The in-house FEM research code MEQSICO is optimized for the solution of large-scale nonlinear EQS problems [10]. The FEM assembly is parallelized on the CPU. When porting the assembly on the GPU, the most time-consuming part is not the calculation of the local contributions but the reduction to a global matrix. Multiple local contributions must be added up for each global matrix entry. Furthermore, the memory needed for all local matrices is much higher than the memory of the global matrix, easily exceeding a single GPU's memory. This requires several cycles of assembly and reduction as well as copying chunks of data between GPUs and host, which is obviously slowing down the overall process.

An assembly of the global stiffness matrix $\mathbf{K}(\mathbf{x})$ is necessary in implicit but not for explicit time-integration schemes. Here, only a vector resulting from a matrix vector multiplication is required. This can be exploited to avoid the time intensive part and only rely on the highly parallelizable part of the assembly.

Here, a GPU based assembly and sparse matrix vector combination is used. The FEM data, nonlinear material parameters and the vectors are handed over to the GPU. Every thread works on a single tetraeder. This differs from other implementations [11], where a whole GPU thread block works on one tetraeder forming the local matrix in shared memory. The increased number of registers per thread in Nvidia's Kepler architecture makes it possible to store all data in thread registers and allow efficient implementations. Finally, after forming the local matrix on the thread, it is directly multiplied with the corresponding vector entries and only these are handed back to the GPUs global memory, decreasing the number of transferred data to less than 10% for second order ansatz functions. The result is directly added to the final right-hand-side vector.

## IV. Multiple Right-Hand side acceleration by PCG Start Vector estimation

As mentioned before, the remaining linear algebra task is a multiple right hand side (MRHS) problem with a constant matrix $\mathbf{M}$. To speed up the procedure, good approximations of the solution vector are beneficial to start the iterative solver.

For example, the Subspace Projection Extrapolation (SPE) scheme [12] generates starting vectors from linear combinations of available preceding solution vectors. Spectral components of the solution are considered which causes the PCG method to converge with respect to the improved effective condition number. A set of m orthonormal vectors is generated from the last (possibly linearly dependent) n solutions by a modified Gram-Schmidt process:

$$\hat{\mathbf{x}}_{j},\, j=1,\ldots,n\xrightarrow{\text{MGS}}\mathbf{V}:=\{\mathbf{v}_{1}\,|\ldots|\,\mathbf{v}_{m\leq n}\} \tag{4}$$

The resulting matrix $\mathbf{V}$ is used to create a projection matrix $\mathbf{V}^{\mathbf{T}}\mathbf{M}\mathbf{V}$ of dimension $m\times m$. As $m\ll\dim(\mathbf{x})$, the solution of this small system can be efficiently calculated with a direct solver. The right-hand-side vector $\mathbf{b}$ is projected with $\mathbf{V}^{\mathbf{T}}$ onto the subspace and solved on the host. The resulting vector is prolonged by multiplication with $\mathbf{V}$ and used as initial vector for the iterative PCG solver

$$\hat{\mathbf{x}}_{m+1,0}:=\mathbf{V}\left[\mathbf{V}^{\mathrm{T}}\mathbf{M}\mathbf{V}\right]^{-1}\mathbf{V}^{\mathrm{T}}\mathbf{b}. \tag{5}$$

Alternatively, the underlying idea of Proper Orthogonal Decomposition (POD) can be used. Recent research [13] aims at replacing the whole system by a reduced order system. In classical POD, many full system solutions ("snapshots") are collected and the projection matrix $\mathbf{V}$ is extracted by singular value decomposition (SVD). Here, the same idea is used but only for the generation of start vectors.

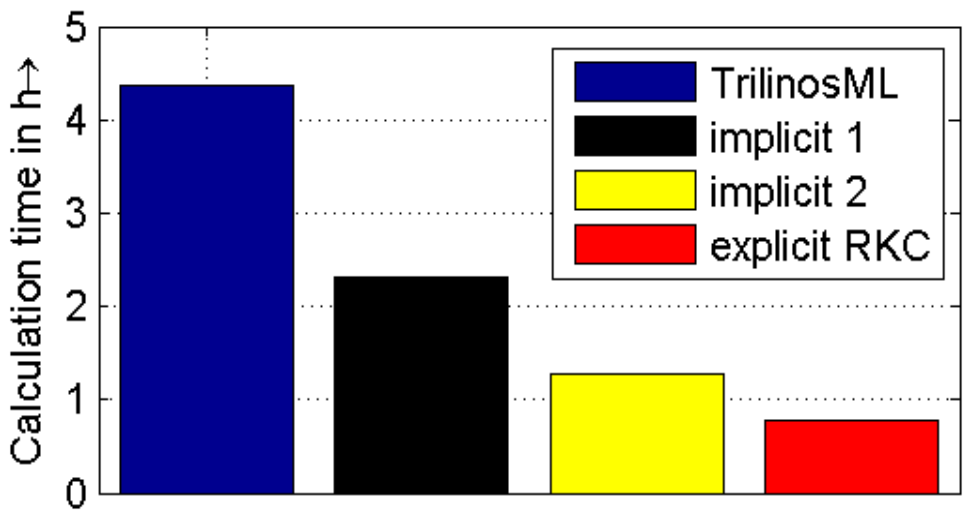

Fig. 2: Calculation time for the EQS-simulation with 1st order ansatz functions using different time integrators and linear solvers

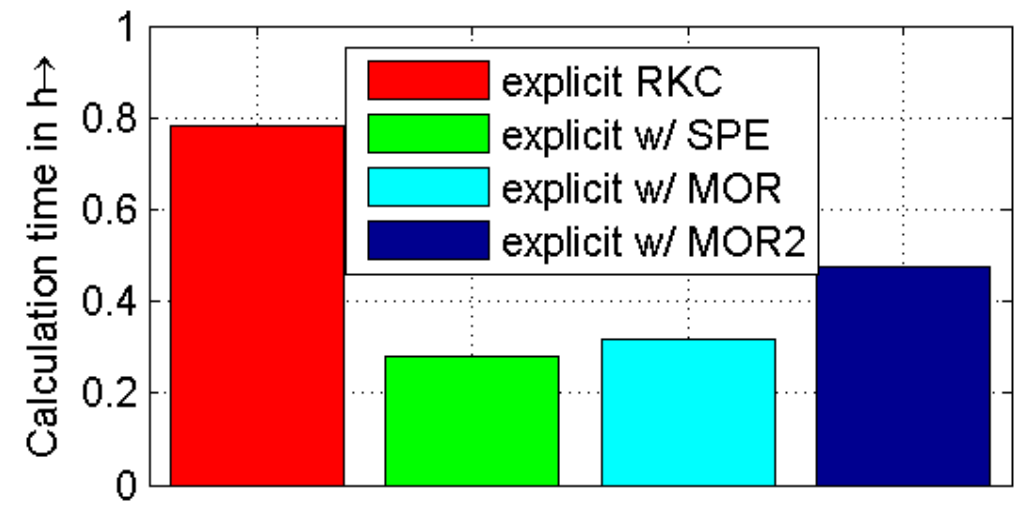

Fig. 3: Calculation time for the EQS-simulation with 1st order ansatz functions using the RKC time integrator and different start value estimators

With this, storing a high number of snapshots leads easily to memory demand that exceeds the available memory capacity. The reduced solution is used as a start vector for the PCG solver. If the start vector is sufficiently accurate, no iterations of the solver are necessary. Otherwise, the PCG iterations will reduce the residuum to the given tolerance.

For this, two approaches are chosen. Like in the conventional POD-MOR method, many snapshots are saved in a snapshot-matrix and decomposed with a SVD into the matrices $\mathbf{V\Sigma U}^T$. Only the $m$ vectors corresponding to the largest singular values are kept. Then the $\dim(\mathbf{x})\times m$ matrix $\mathbf{V}$ is used to generate the projection matrix like (5). In the first approach, many snapshots are taken and kept for the rest of the simulation. Thus, the time-consuming calculation of the reduced system in (5) is performed only once.

In a second approach, the POD-MOR is used like the SPE initial value estimator, considering the last solutions of the simulation process. New values are added to the snapshot matrix if a given number of PCG iterations is exceeded. If the maximum number of snapshots is reached, the oldest ones are overwritten. This means that the set of solutions is changing and the reduced system has to be calculated multiple times.

## V. Numerical Example

As a numerical example, the electrical field of high voltage insulator with a microvaristor electric field grading coating of 1 mm thickness is calculated. With 1st order FEM ansatz functions the model consists of 2.1 mio. degrees of freedom (DoF) and 30.4 mio. non-zero matrix entries (nnz). Using 2nd order ansatz functions leads to 17.4 mio. DoF and 494 mio. nnz. The implicit time integrator SDIRK3(2) [2] is used as the standard time integrator for the simulation code MEQSICO [10]. Alternatively, the proposed explicit RKC time integrator is employed [9]. The local time-integration error is set to values below $1\times10^{-2}$. The maximum tolerance of the relative residuum of the linear solver is set to $1\times10^{-12}$ and the maximum nonlinear residuum of the Newton-Raphson-iteration scheme is set to $1\times10^{-8}$. The GPU compute server is equipped with two Intel Xeon 2660 CPUs with a total of 20 cores and 4 Nvidia Tesla K20 GPUs using Cuda 7.5, Thrust 1.8, CUSP 0.5.1 [14]. The code uses a CPU based state-of-the-art solver library Trilinos ML [15] implemented as the standard linear solver for MEQSICO. Due to its memory requirements the model requires at least four Tesla K20 GPUs when solved with 2nd order FEM ansatz functions.

TABLE I
Time required for the assembly of the conductivity Matrix and the linear system Setup

| Shape order | Time integrator | **K**-Assembly in [s] | RHS and matrix setup [s] |
|---|---|---|---|
| 1st order | SDIRK32 | 1,8 | 2,7 |
| 1st order | RKC | 0,013 | 0,02 |
| 2nd order | SDIRK32 | 28,7 | 62,0 |
| 2nd order | RKC | 0,2 | 0,3 |

Table I shows, the amount of time spend on initial operations of each stage. These are at first the assembly of the nonlinear matrix and the total time for setting up the linear system. This includes the calculation of the RHS vector as well as the addition of the mass and stiffness matrix in the implicit time-integration scheme. The combined assembly and multiplication described in Sec. III outperforms the classical assembly by over two orders of magnitude. The overall setup of the linear system achieves an even higher speedup for the explicit approach compared to the implicit time-integration scheme by avoiding the costly addition of two large matrices.

Fig. 2 compares the time spend on solving the given problem with first order ansatz functions. The initial offset for setting up the system from FEM grid coordinates is neglected, as it is the same for all variants.

In Fig. 2, the label “Trilinos ML” refers to the code version using the host based linear PCG solver with Trilinos ML providing the AMG preconditioner. Due to the changing Jacobian matrix, the preconditioner is set up in every time step. “Implicit 1” uses the Multi-GPU AMG-CG solver as presented in [8]. “Implicit 2” uses the before mentioned Multi-GPU solver and the adaptive AMG preconditioner presented in [16]. “Explicit RKC” addresses the explicit approach described in this paper. Here, the explicit approach outperforms all implicit approaches with speedup factors 5.6 vs. Trilinos ML, 3.0 vs. “Implicit 1” and 1.636 vs. “Implicit 2”, although the number of time steps is twice as high as for the implicit approach.

Fig. 3 shows the presented explicit approach combined with start value estimators. Due to the resulting reduction of PCG iterations a speedup of factor 2.8 for the SPE scheme and a factor of 2.5 for the “MOR” scheme calculation with only one fixed snapshot matrix can be achieved. The “MOR2” scheme, which repeatedly calculates reduced systems from the last solutions, is slower than the other start value estimators. Even though it needs the least number of overall CG iterations a significant amount of time is lost due to multiple SVDs and calculations of reduced systems.

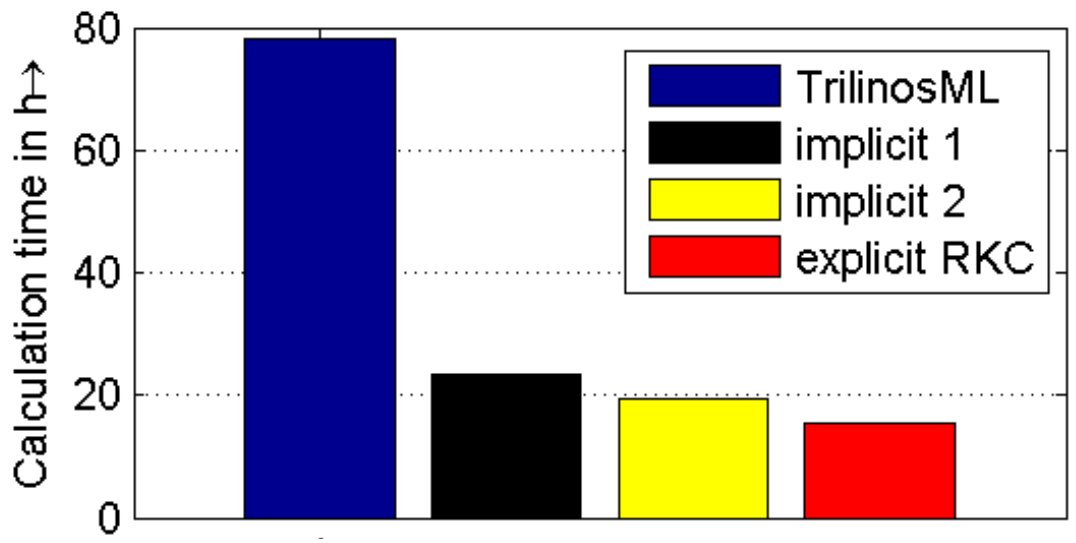

Fig. 4: Calculation time for the EQS-simulation with 2ndt order ansatz functions using different time integrators and linear solvers

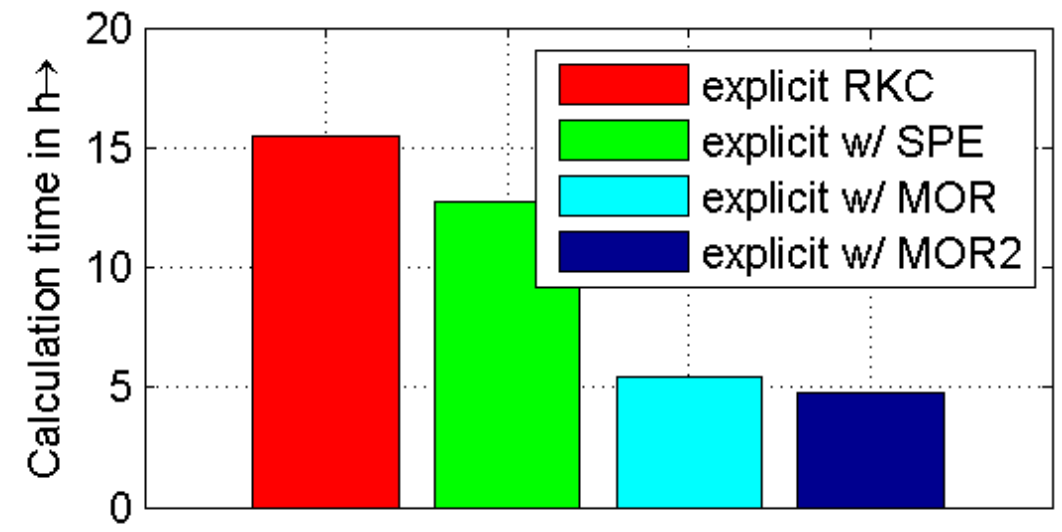

Fig. 5: Calculation time for the EQS-simulation with 2nd order ansatz functions using the RKC time integrator and different start value estimators

Fig. 4 shows the analogue timings as in Fig. 2 for 2nd order FEM ansatz functions and the general behavior of the solvers remains the same. The main difference is the higher speedup between CPU and GPU based calculations. Here the GPUs are better utilized because there are more entries per row in the system matrix and thus more time is spend solving the linear system.

Fig. 5 shows the timings as in Fig. 3 for the 2nd order FEM ansatz functions. The situation differs from the one for the 1st order FEM ansatz functions. Here, the SPE start vector generation is not able to estimate the result vector with the same quality as for the first order case. This is due to the higher number of matrix entries and complexity causing a much higher number of PCG iterations. The POD-MOR gives better estimations. As iterations are more time consuming for 2nd order FEM ansatz functions, a low number of PCG iterations is key to a fast solution process. In the "MOR2" start vector generation the effort for performing multiple SVDs is overcompensated due to the reduced number of PCG iterations resulting in the fastest solution process. While the "MOR" scheme needs an overall number of 10,045 PCG iterations, "MOR2" results in 5,651 iterations.

## VI. Conclusion

An explicit time-integration scheme for nonlinear electro-quasistatic field problems accelerated by multiple GPUs was presented. Beside the multi-GPU accelerated AMG-PCG solver, a combined GPU based FEM assembly and sparse matrix vector product accelerated the assembly process for the nonlinear conductivity matrix. With these accelerations, the explicit time-integration scheme outperforms implicit time-integration schemes significantly, even though more time steps are needed. For the constant system matrix MRHS problem within the explicit scheme further accelerations were achieved by using SPE or POD-MOR based start value estimators.

## VII. Acknowledgements

Part of this work was funded under grants CL 143/10-2 and SCHO-1562/1-1 by the Deutsche Forschungsgemeinschaft (DFG) and supported by the German 'Excellence Initiative' and the Graduate School CE TU Darmstadt.